\documentclass[a4paper, times, 10pt,twocolumn]{article}
\usepackage[top=4.9cm,bottom=3.7cm,left=1.5cm,right=1.5cm]{geometry}
\usepackage{ICMLC}
\usepackage{times}
\usepackage{graphicx}
\usepackage{indentfirst}
\usepackage{latexsym}
\usepackage{microtype}
\usepackage{booktabs}
\usepackage{amsmath}
\usepackage{amsfonts}
\usepackage[numbers,sort&compress]{natbib}
\usepackage[table]{xcolor}
\usepackage{url}
\usepackage{color}
\usepackage{hyperref}
\usepackage{cleveref}
\usepackage{pifont}
\usepackage[margin=8pt,font=footnotesize,labelfont=bf,labelsep=period]{caption}

\pdfpagewidth=\paperwidth
\pdfpageheight=\paperheight
\newcommand{\cmark}{\ding{51}}
\definecolor{smallcolor}{HTML}{D8BFD8}
\definecolor{mediumcolor}{HTML}{9370DB}
\definecolor{largecolor}{HTML}{4B0082}

\begin{document}

\title{Building Specialized Software-Assistant ChatBot with Graph-Based Retrieval-Augmented Generation}

\author{
\bf{Mohammed Hilel$^{1*}$} \and
\bf{Yannis Karmim$^{1*}$} \and
\bf{Jean De Bodinat$^{1}$} \and
\bf{Réda Sarehane$^{2}$} \and
\bf{Antoine Gillon$^{2}$}
\\
$^{1}$RAKAM AI, Paris, France
\quad
$^{2}$Lemon Learning, Paris, France
\\
$^{*}$Equal contribution
}

\maketitle \thispagestyle{empty}

\begin{abstract}
Digital Adoption Platforms (DAPs) support employees in the use of complex enterprise software, but creating and maintaining in-app guidance remains largely manual. We propose an industrial Graph-based Retrieval-Augmented Generation (Graph-RAG) pipeline that automatically converts web applications into \emph{state-action} knowledge graphs and retrieves a compact, connected subgraph relevant to a user query. The retrieved subgraph is textualized and injected into the LLM prompt to generate grounded step-by-step instructions without retraining, making it compatible with black-box APIs. We also introduce \texttt{SoftwareQA}, a benchmark of $\approx$1200 multiple-choice questions across 4 enterprise applications. Experiments show consistent accuracy gains across model sizes, reaching up to +16 points with graph augmentation. Code and dataset will be release. 
\end{abstract}

\vspace{-0.35cm}
\Section{Introduction}
\label{sec:intro}
\vspace{-0.35cm}
Digital Adoption Platforms (DAPs) help employees navigate complex business software such as CRM~\cite{NutshellCRM2023}, ERP~\cite{NetSuiteERP2023}, and HRMS~\cite{HRMorningATS2023} systems, but authoring their step-by-step guides remains largely manual~\cite{Latka2024}. LLMs~\cite{Brown2020gpt3,llama3} could automate this, yet they hallucinate without structured knowledge of the target application~\cite{huang2025survey}, and fine-tuning is impractical due to catastrophic forgetting~\cite{gekhman_does_2024} and incompatibility with black-box APIs~\cite{openai2023gpt4}. Retrieval-Augmented Generation (RAG)~\cite{lewis2021retrievalaugmentedgenerationknowledgeintensivenlp} currently outperforms fine-tuning for knowledge injection~\cite{ovadia_fine-tuning_2024}, and graph-structured retrieval further improves precision~\cite{Xu_2024}.\textbf{ However, standard RAG assumes the availability of well-structured and maintained textual documentation}. In enterprise settings, such documentation is sometime incomplete, outdated, or entirely unavailable. Moreover, raw document retrieval does not capture the procedural and state-dependent nature of UI interactions.

We present a Graph-RAG framework that addresses this gap. Our contributions are: (i) a \textbf{Software-to-Graph pipeline} that automatically extracts state-action graphs directly from web interfaces, removing the dependency on pre-existing documentation; (ii) a \textbf{graph retrieval} method adapted to large state-action graphs and compatible with black-box LLMs; (iii) \textbf{\texttt{SoftwareQA}}, a benchmark across four enterprise applications.
\begin{figure}[!ht]
    \centering
    \includegraphics[width=\linewidth]{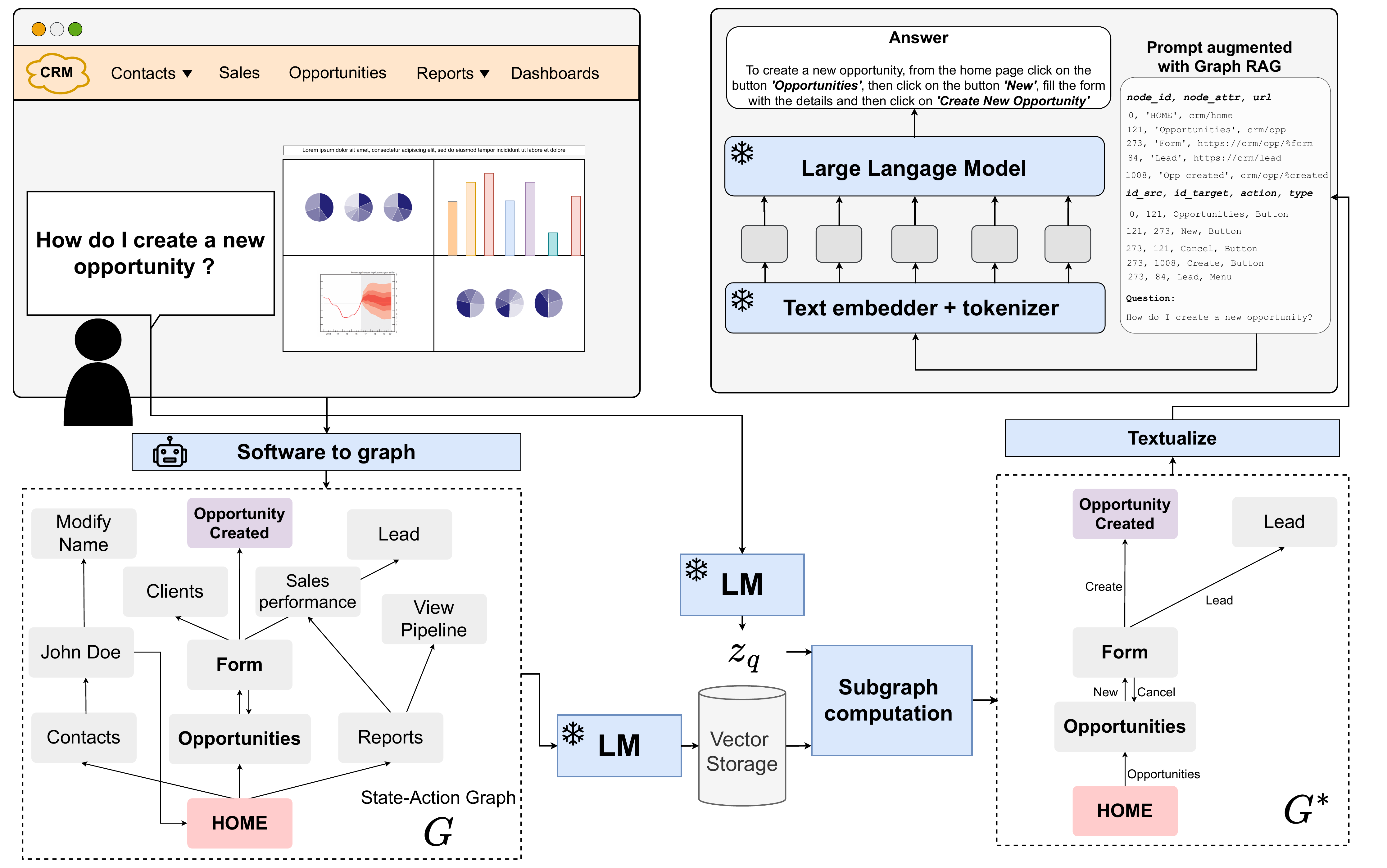}
    \caption{Graph-RAG overview: (1) the software is converted into a state--action graph $G$; (2) the user query is embedded to score graph elements; (3) a compact subgraph $G^*$ is extracted via PCST; (4) $G^*$ is textualized and provided as context to the LLM.}
    \label{fig:introduction}
\end{figure}
\vspace{-0.2cm}
\section{Method}
\label{sec:design}
\vspace{-0.35cm}
\textbf{Software-to-Graph.} We define a knowledge graph $G=(V,E)$ where nodes $u \in V$ are interface states and edges $e \in E$ are user actions with textual attributes $x_u$, $x_e$. A Playwright-based bot crawls the web application via breadth-first traversal, detecting clickable elements through DOM inspection. The pipeline requires only login credentials and a home page URL, is domain-agnostic, and produces graphs updated incrementally as the software evolves. Table~\ref{tab:stat_graph} reports graph statistics for our four applications.

\begin{table}[t]
\caption{Extracted knowledge graphs statistics.}
\label{tab:stat_graph}
\vskip 0.1in
\begin{center}
\begin{small}
\begin{sc}
\begin{tabular}{lrrrr}
\toprule
Software & Type & \#Nodes & \#Edges & Time(s) \\
\midrule
SalesForce & CRM & 7640 & 7655 & 5217 \\
Dolibarr   & ERP & 1714 & 2196 & 9652 \\
Pipedrive  & CRM &  120 &  694 &  670 \\
Ivalua     & ERP &  450 &  464 &  449 \\
\bottomrule
\end{tabular}
\end{sc}
\end{small}
\end{center}
\vskip -0.1in
\end{table}

\textbf{Graph-based retrieval.} Since full graphs are too large for an LLM prompt, we apply the Prize-Collecting Steiner Tree (PCST) algorithm~\cite{bienstock1993note} to extract a minimal connected subgraph $G^*$. Node and edge descriptions are encoded with a multilingual Sentence-BERT~\cite{Reimers2019SentenceBERTSE} (\texttt{MiniLM-L12-v2}); cosine similarity with the query selects top-$k$ candidates. PCST then maximizes cumulative relevance prizes while minimizing traversal cost, enforcing a single connected component (\texttt{pcst\_fast}, \texttt{num\_cluster=1}). The user's current UI state is pinned with similarity score 1, anchoring the subgraph to the user's context.

\textbf{Prompt construction.} The textualized subgraph is concatenated with the user query and fed to a frozen LLM. Listing edges as \texttt{(src, tgt, action)} enables multi-hop reasoning~\cite{zhang2024can} and reduces hallucinations~\cite{Xu_2024}. Unlike G-Retriever, we omit GNN and MLP projection layers, making our approach directly compatible with any black-box LLM API without retraining.
\vspace{-0.35cm}
\section{Experiments}
\vspace{-0.35cm}
\textbf{\texttt{SoftwareQA}.} Questions and distractors were generated from official documentation using \texttt{gpt-oss-120b} in a three-pass pipeline (question generation $\to$ distractor generation $\to$ distractor filtering). This design also validates that our KG covers the official documentation, a useful property when docs are sparse or outdated. The benchmark covers three skills: \textit{navigation}, \textit{configuration}, and \textit{concept} understanding, with $\approx$300 questions per application (Table~\ref{tab:dts_stat}).

\begin{table}[ht]
\centering
\caption{MCQ example from \textsc{SoftwareQA}.}
\label{tab:mcq-example}
\small
\renewcommand{\arraystretch}{1.2}
\begin{tabular}{@{}lp{6cm}@{}}
\toprule
\textbf{Field} & \textbf{Content} \\
\midrule
ID / Type & 285 / Configuration \\
Question & Navigation path to configure VAT accounts in Dolibarr? \\
A\,(\cmark) & Accounting $\to$ Configuration $\to$ VAT $\to$ Add rate \\
B & Home $\to$ Settings $\to$ Modules $\to$ Enable VAT \\
C & Third Parties $\to$ Configuration $\to$ Taxes $\to$ Create rate \\
D & Bank $\to$ Accounts $\to$ VAT tab $\to$ Add rate \\
\bottomrule
\end{tabular}
\end{table}

\begin{table}[htbp]
    \centering
    \caption{\texttt{SoftwareQA} dataset statistics.}
    \label{tab:dts_stat}
    \setlength{\tabcolsep}{3pt}
    \begin{tabular}{lcccc}
    \toprule
    Software & Nav. & Config. & Concept & Total \\
    \midrule
    Salesforce & 191 & 103 & 6   & 300 \\
    Dolibarr   & 134 & 159 & 7   & 300 \\
    Pipedrive  & 244 &  27 & 28  & 299 \\
    Ivalua     &  77 &  46 & 174 & 297 \\
    \bottomrule
    \end{tabular}
\end{table}

\textbf{Results.} Accuracy is measured as the fraction of correctly selected choices, making results fully reproducible. As shown in Table~\ref{tab:overall-accuracy}, Graph-RAG improves every model on every dataset (+0.2 to +16.4 pts). Smaller models benefit the most: Ministral-Small +GRAG reaches the unaugmented Qwen3-480B level, showing graph retrieval can substitute for model scale. Gains are highest on Ivalua, whose domain-specific procurement vocabulary is harder for LLMs to internalize.
\vspace{-0.15cm}
\begin{table}[htbp]
\small
\centering
\caption{Accuracy (\%). \textbf{SF}: Salesforce, \textbf{DB}: Dolibarr, \textbf{PD}: Pipedrive, \textbf{IV}: Ivalua.}
\label{tab:overall-accuracy}
\setlength{\tabcolsep}{4pt}
\begin{tabular}{l@{\hspace{4pt}}cccc}
    \toprule
    \textbf{Model} & \textbf{SF} & \textbf{DB} & \textbf{PD} & \textbf{IV} \\
    \midrule
    \multicolumn{5}{l}{\textit{\textcolor{smallcolor}{\textbf{Small}}}} \\
    Ministral-S        & 75.6 & 79.0 & 94.9 & 77.9 \\
    Ministral-S +GRAG  & \textbf{83.0} & \textbf{83.3} & \textbf{95.1} & \textbf{94.3} \\
    \midrule
    \multicolumn{5}{l}{\textit{\textcolor{mediumcolor}{\textbf{Medium}}}} \\
    Mistral-M          & 79.7 & 78.0 & 91.6 & 76.3 \\
    Mistral-M +GRAG    & \textbf{87.0} & \textbf{84.3} & \textbf{94.6} & \textbf{95.3} \\
    \midrule
    \multicolumn{5}{l}{\textit{\textcolor{largecolor}{\textbf{Large}}}} \\
    Qwen3-480B         & 79.3 & 83.7 & 96.0 & 81.9 \\
    Qwen3-480B +GRAG   & \textbf{87.0} & \textbf{86.3} & \textbf{96.3} & \textbf{94.3 }\\
    \bottomrule
\end{tabular}
\end{table}

\section{Conclusion and industrial impact}
\vspace{-0.35cm}
We presented a Graph-RAG framework that converts enterprise web applications into state--action knowledge graphs to ground LLM-generated software assistance. Compatible with black-box APIs and requiring no fine-tuning, it improves accuracy on \texttt{SoftwareQA}. Future work includes more robust graph extraction and user-level personalization. Our approach reduces manual authoring and maintenance effort in Digital Adoption Platforms, shortens adaptation cycles when enterprise software evolves, and lowers operational costs by scaling guidance generation across applications. Beyond assistance, grounding LLMs in executable UI knowledge also opens the path toward LLM-driven UI agents for partial enterprise workflow automation.
\bibliographystyle{IEEEtran}
\bibliography{main}

\end{document}